\documentclass[aps,nofootinbib,twocolumn]{revtex4}

\usepackage{graphicx}
\usepackage[dvips]{color}
\usepackage{dcolumn}
\usepackage{bm}
\usepackage{color}

\begin{document}

\title{Electronic structure and ionicity of actinide oxides from first principles calculations}

\author{L. Petit}
 \email{lpetit@phys.au.dk}
 \affiliation{Department of Physics and Astronomy, Aarhus University, DK-8000 Aarhus C, Denmark}
\author{A. Svane}
 \affiliation{Department of Physics and Astronomy, Aarhus University, DK-8000 Aarhus C, Denmark}
\author{Z. Szotek}
 \affiliation{ Daresbury Laboratory, Daresbury, Warrington WA4 4AD, UK }
\author{W. M. Temmerman}
 \affiliation{ Daresbury Laboratory, Daresbury, Warrington WA4 4AD, UK }
\author{G. M. Stocks}
 \affiliation{ Materials Science and Technology Division, Oak Ridge National Laboratory, Oak Ridge, Tennessee 37831, USA }

\date{\today}
\begin{abstract}
The ground state electronic structures of the actinide oxides AO, A$_2$O$_3$ and AO$_2$ (A=U, Np, Pu, Am, Cm, Bk, Cf)
are determined
from first-principles calculations, using the self-interaction corrected local spin-density (SIC-LSD) approximation.
Emphasis is put on the degree of f-electron localization, which for AO$_2$ and A$_2$O$_3$
is found to follow the stoichiometry, namely corresponding to A$^{4+}$ ions in the dioxide and A$^{3+}$ ions in the
sesquioxides. In contrast,
the A$^{2+}$ ionic configuration is not favorable in the monoxides, which therefore become metallic.
The energetics of the oxidation and reduction of the actinide dioxides is discussed, and it is found that
the dioxide is the most stable oxide for the actinides from Np onwards.
Our study reveals a strong link between preferred oxidation number
and degree of localization which is confirmed by comparing to the ground state configurations of the corresponding
lanthanide oxides. The ionic nature of the actinide oxides emerges from the fact that only those compounds will form where
the calculated ground state valency agrees with the nominal
valency expected from a simple charge counting.

\end{abstract}

\maketitle

\section{Introduction}

Actinide oxides play a dominant role in the nuclear fuel cycle.~\cite{Choppin} For many years, uranium dioxide has been the main fuel component
in commercial nuclear reactors. The "burning" of UO$_2$ results in considerable amounts
of Np and Pu isotopes, as well as smaller quantities of
minor actinides such as Am, Cm, Bk and Cf. In original, "once through" reactors, the highly radioactive waste that is 
produced  resulted in very troublesome long time storage requirements; particularly of Pu. However, it was soon realized that the Pu could be reprocessed from the spent fuel
and used as alternative fuel in a new generation of reactors.
The environmental and energy production issues apart, the fact that Pu is obtained from the decommisioning of
nuclear weapons is yet another important consideration concerning its use as nuclear fuel.
A mixture of UO$_2$ and PuO$_2$, the so-called mixed oxide (MOX), where Pu is blended with either natural
or depleted Uranium,
constitutes the preferred, Pu containing, fuel in existing nuclear reactors.
Lately, Np, as well as the minor actinides that accumulate during nuclear reactor operation,
are also being considered for reprocessing.~\cite{Konings} Once separated from the spent fuel, these actinides can
either be incorporated
in durable ceramic waste for safe long time storage (and possible later recovery) or transmuted
from long-lived isotopes to less radiotoxic short-lived isotopes through irradiation, thus taking part in the fuel cycle and reducing the long term
nuclear waste management problem. Again oxides are being considered both with respect to the materials being used as fission/transmutation
targets (AO$_2$)~\cite{Beauvy}
and for the direct storage in the shape of durable ceramic glasses (AO$_2$, A$_2$O$_3$).~\cite{Konings}
For example PuO$_2$ has long been the compound of choice for depositing Pu in long time repositories,
given the observed stability with respect to oxidation.~\cite{Weigel}

Regarding their behaviour under both reactor operation and storage conditions, it is crucial to understand
the thermochemistry, thermophysics, and materials science of the actinide compounds. Given the toxicity of the
materials involved, computer simulations, such as thermodynamic modelling~\cite{Gueneau} or, as in the present paper,
electronic structure calculations can provide fundamental insights at a level not achieveable through experiment alone.
Here we wish to focus specifically on the $f$-electrons, their contribution to the groundstate electronic properties 
of the actinide oxides, the role they play with regards to stability towards oxidation, and their behaviour under
ionic bonding conditions. 

When modelling the electronic structure of actinide materials, the most distinguishing feature is the increasing importance of correlations
across the series from U to Cf, as the nature of the $f$-electrons
changes from delocalized in the early actinides to localized in the later actinides.~\cite{Johansson_AcMetal,Svane_AcMetal}
Electronic structure calculations, based on the local spin density approximation (LSDA), do not
take into account strong on-site correlations beyond the homogeneous electron gas, and therefore can not adequately describe the localized phase of
actinide materials. Thus the LSDA,~\cite{Maehira} or even the generalized gradient
approximation (GGA)~\cite{Boettger_PuO2} (which extends beyond the LSDA by taking into account charge
density gradients), wrongly predicts a metallic ground state for UO$_2$, PuO$_2$ and Pu$_2$O$_3$, although the equilibrium lattice parameters and cohesive properties
are found to be in rather good agreement with experiment.~\cite{Freyss,Boettger_PuO2,Jomard}

A number of schemes have been developed that augment the standard bandstructure framework to include the effects
of strong correlations on the electronic structure.
In the LDA+U approach~\cite{Anisimov} an effective Coulomb parameter $U$ is introduced that separates the $f$-manifold
into the upper and lower Hubbard bands and removes $f$-degrees
of freedom from the Fermi level. The more advanced dynamical mean-field theory (DMFT) approach provides a description of the competing trends towards localization on the
one hand and itineracy on the other
hand by taking into account the local quantum fluctuations
missing in the static LDA+U treatment ~\cite{Kotliar_PhysicsToday, Savrasov_Pu_Nature}; albeit still at the cost of the introduction of the  $U$-parameter. 
The hybrid density functional~\cite{Becke} theory
implements an exchange-correlation functional where a fraction of the exact non-local
exchange interaction from Hartree-Fock theory is mixed with the local or semi-local exchange energy of LSDA or GGA with the result that the troublesome effects of the known self-interaction error present in the standard LSDA and GGA calculations are reduced.

The self-interaction
corrected (SIC)-LSD approach~\cite{Perdew_Zunger} used in the current work removes
the self-interaction error that occurs in the LSDA, thereby leading to an improved description of the static Coulomb correlations
of the f-electrons. The self-interaction correction associates  an energy gain with electron localization,
which competes with the opposing trend of band formation,
providing a dual picture of combined localized and band like $f$-electrons. The method is fully {\it ab-initio} as
both kinds of electrons are treated on an equal footing, with no adjustable parameters. A comparative study of MnO,
involving SIC-LSD, LDA+U, and the hybrid functional methods,
was published by Kasinathan {\it et al.}.~\cite{Kasinathan}

The present paper is organized as follows. In the following section, we give a short introduction to the SIC-LSD
band structure method. In section III, we present our SIC-LSD results for the ground state properties of (A)
the monoxides, (B) the sesquioxides, and (C) the dioxides where we also consider oxidation/reduction energies.
In section IV, we give a summarizing discussion of the results,
also concerning the relation between $f$-electron localization and oxidation by comparing to the lanthanide oxides.
The conclusion of our paper is presented in section V.

\section{The SIC-LSD methodology}

The LSD approximation to the exchange and correlation energy introduces an unphysical interaction of an
electron with itself~\cite{Perdew_Zunger} which, although insignificant for extended band states, may lead to uncontrollable
errors in the description of atomic-like localized states, for example the $f$-electrons in the later actinides.
The SIC-LSD method~\cite{Temmerman_LNP,Svane_SIC} corrects for this spurious
self-interaction
by adding to the LSD total energy functional an explicit energy contribution for an electron to localize.
The resulting, orbital dependent, SIC-LSD total energy functional has the form
\begin{equation}
\label{Esic}
E^{SIC-LSD}=E^{LSD}+E_{so}-\Delta E_{sic},\\
\end{equation}
where
\begin{eqnarray}
\label{Esicdetail}
\hspace{-4mm}
E^{LSD}&=&\sum_{\alpha }^{occ.}\langle \psi_{\alpha} | -\nabla^2 | \psi_{\alpha} \rangle+U[n]+V_{ext}[n] \\
       &+&E_{xc}^{LSD}[n_{\uparrow},n_{\downarrow}], \nonumber \\
E_{so}&=&\sum_{\alpha }^{occ.}\langle \psi_{\alpha} | \xi(\vec{r})\vec{l}\cdot\vec{s} | \psi_{\alpha} \rangle, \\
\Delta E_{sic}&=&\sum_{\alpha }^{occ.}\delta _{\alpha }^{SIC}=\sum_{\alpha }^{occ.} \left \{ U[n_{\alpha}]+E_{xc}^{LSD}[n_{\alpha}] \right \}.
\end{eqnarray}
Here the sums run over all occupied electron states $\psi_{\alpha}$.
As usual, the LSD total energy functional (2)
is decomposed into the kinetic energy, the Hartree energy, the
interaction energy with the atomic ions, and the exchange and
correlation energy.
The spin-orbit interaction (3) couples the band Hamiltonian for the spin-up and spin-down channels,
{\it i.e.}  a double secular problem must be solved.
The spin-orbit parameter,
\[
\xi(r)=-\frac{2}{c^2}\frac{dV}{dr},
\]
 in atomic Rydberg units,
is calculated from the self-consistent potential.
The self-interaction energy (4) consists of the self-Coulomb and self-exchange-correlation energies of the occupied orbitals $\psi_{\alpha}$
with the orbital charge density $n_{\alpha}$.

For itinerant states, the self-interaction $\delta _{\alpha }^{SIC}$ vanishes identically, while
for localized (atomic-like) states $\delta _{\alpha }^{SIC}$  may be appreciable.
Thus, the self-interaction correction constitutes a negative energy
contribution gained by an electron upon localization,
which competes with the band formation energy gained by the electron
if allowed to delocalize and hybridize with the available conduction states.
Different localized/delocalized configurations are realized by assuming different numbers of localized
states - here $f$-states on actinide-atom sites. For $s$- and $p$-states, $\delta _{\alpha }^{SIC}$ is
never competitive compared to the corresponding gain in band formation energy, and turns out to be positive.
Since the different localization scenarios constitute distinct
local minima of the same energy functional, $E^{SIC-LSD}$, their total energies may be compared and
the global energy minimum then defines the ground state total energy {\em and} the valence configuration
of the actinide-ion. This latter is defined as the integer number of electrons available for band formation,
\begin{equation}
N_{val}=Z-N_{core}-N_{SIC},
\end{equation}
where $Z$ is the atomic number, $N_{core}$ is the number of atomic core electrons, and $N_{SIC}$ is the
number of SIC-localized $f$-electrons.
In the remainder of this paper we will be using two interchangeable nomenclatures, $f^n$ and A$^{m+}$, to describe the configuration of the
actinide ion, implying $n=N_{SIC}$ and $m=N_{val}$, respectively. The total number of $f$-electrons may be larger
than $n$, since, in addition to the
$n$ localized $f$-states, the band states also contribute to the total $f$-count on a given ion. Note that our calculated valencies refer
to the number of actinide electrons that contribute to bonding, and thus do
not necessarily coincide with the nominal (ionic) valency of a compound: For PuO$_2$ for example, the Pu$^{4+}$ would agree with 
an ionic picture, while the Pu$^{3+}$, and Pu$^{5+}$ valency configurations would indicate more covalent behaviour.

The SIC-LSD approach is fully ab-initio,
as both localized and delocalized states are expanded in the same set of basis functions
and are, thus, treated on an equal footing.
If no localized states are assumed, $E^{SIC-LSD}$ coincides with the
conventional LSD functional, {\it i.e.}, the Kohn-Sham minimum of the $E^{LSD}$
functional is also a local minimum of $E^{SIC-LSD}$.

Given the total energy functional $E^{SIC-LSD}$, the computational procedure is as for the LSD case,
{\it i.e.} minimization is accomplished by iteration until self-consistency.
In the present work, the electron wavefunctions are expanded in
the linear-muffin-tin-orbital (LMTO) basis functions.\cite{Andersen_LMTO}
The atomic spheres approximation (ASA) is used, whereby the crystal volume is divided into
slightly overlapping atom-centered spheres of a total volume equal to the actual volume.
A known shortcoming of the ASA is that different crystal structures have different degrees of overlap of the ASA 
spheres resulting in substantial {\it relative} errors in the evaluation of the total energy. While this inhibits
the comparison of energies of different crystal structures, when comparing 
the energies of different localization scenarios within the same crystal
structure, the ASA error is of minor influence. To improve the packing of the structure
empty spheres have been introduced on high symmetry interstitial sites.

\section{Results}

\subsection{Actinide Monoxides}

\subsubsection{Background information}

There exists to date no convincing evidence that actinide oxides can form in the 1:1 stoichiometry. The experimental lattice parameters
that we cite in Table \ref{monoxidesTable} come from early reports on these compounds, and have so far not been confirmed.
There have been no claims of bulk UO having ever been synthesized, and reports of an UO surface phase on U metal for
low exposures to O,\cite{Allen,Ellis} and UO thin films~\cite{Lam} could not be reproduced.\cite{Winer}
It has been suggested that the observed thin films actually represent uranium oxynitrides (UN$_x$O$_{1-x}$) and oxycarbides (UC$_x$O$_{1-x}$), that form
in the presence of N$_2$ or C, and at low oxygen pressure.\cite{Eckle}
It has similarly been concluded that neither bulk NpO, nor a corresponding NpO surface phase will form.\cite{Richter,Naegele}
Preparation of PuO and AmO,~\cite{Akimoto} as well as possibly BkO,~\cite{Fahey} has been neither substantiated, nor has it been dismissed.
\begin{table}
\caption{Actinide monoxide data. Column 2: Groundstate configuration. Column 3: Energy
difference in eV between the ground state and the ideal ionic divalent configuration. Column 4: Groundstate density of states at the Fermi level
(in units of states per eV and formula unit).
Column 5: Calculated lattice parameters $a_0^{calc}$ (in {\AA}) . Column 6: Experimental lattice parameter $a_0^{exp}$ (in {\AA}), where known (measurements on UO, and NpO refer to thin film data,
measurements on PuO, AmO, and BkO refer to bulk data).}
\begin{ruledtabular}
\begin{tabular}{|c|c|c|c|c|c|}
AO   & Config. & E$_{GS}$-E$_{II}$  & n$(E_F) $ & a$_0^{calc}$  & a$_0^{exp}$  \\
\hline
  UO       & f$^1$ (U$^{5+}$) & -1.93  & 5.8 & 4.94 & 4.92$^a$ \\
\hline
  NpO      & f$^3$ (Np$^{4+}$) & -1.73  & 1.1 & 4.99 & 5.01$^b$ \\
\hline
  PuO      & f$^5$ (Pu$^{3+}$) & -0.58  & 5.9 & 5.13 & 4.960$^c$ \\
\hline
  AmO      & f$^6$ (Am$^{3+}$) & -0.14  & 7.3 & 5.14 & 5.045$^c$ \\
\hline
  CmO      & f$^7$ (Cm$^{3+}$) & -1.14  & 2.6 & 5.01 & - \\
\hline
  BkO      & f$^8$ (Bk$^{3+}$) & -0.65  & 1.4 & 4.97 & 4.964$^d$ \\
\hline
  CfO      & f$^9$ (Cf$^{3+}$) & -0.20  & 9.5 & 4.97 & - \\
\hline
  EsO      & f$^{10}$ (Es$^{3+}$) & 0.00 & 4.0 & 4.92 & - \\
\hline
  EsO      & f$^{11}$ (Es$^{2+}$) & 0.00 & 0.0 & 5.06 & - \\
\end{tabular}
\end{ruledtabular}
$^a$Reference \onlinecite{Lam}
$^b$Reference \onlinecite{Pearson}
$^c$Reference \onlinecite{Akimoto}
$^d$Reference \onlinecite{Fahey}
\label{monoxidesTable}
\end{table}

\subsubsection{SIC-LSD electronic structure}

We have calculated the electronic structure of the
monoxides with the SIC-LSD method in order to establish the ground state properties for the hypothetical NaCl structure. Ferromagnetic
arrangement of the spins has been assumed in these calculations. The results are summarized in Table \ref{monoxidesTable}. We find the
trivalent configuration to be energetically most favourable for all the monoxides, except UO and NpO that
respectively prefer the U$^{5+}$ and Np$^{4+}$ ground state configurations.
\begin{figure}
\begin{center}
\includegraphics[width=70mm,clip,angle=-90]{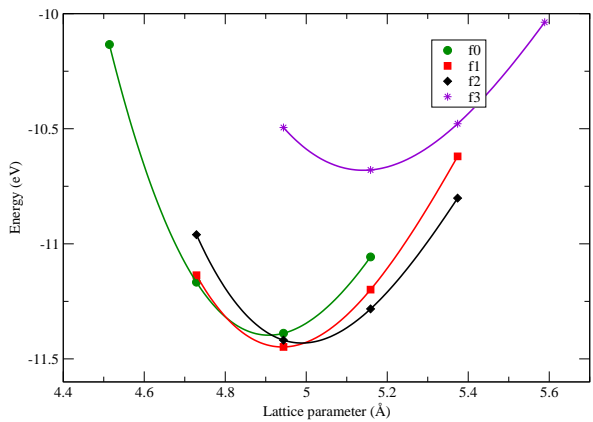}
\caption{
\label{UOenergy}
Total energy of UO in the rocksalt structure assuming different localized/delocalized $f$-electron scenarios. 
}
\end{center}
\end{figure}
Concerning UO, it should be noted
here that even though we find the U $f^1$ configuration to be the ground state, this energy scenario is close to degenerate
with the fully itinerant LSD ($f^0$) and $f^2$ scenarios, as shown in figure \ref{UOenergy}. This indicates that
the $f$-electron manifold lingers between the localized and delocalized pictures, i.e. correlations are strong,
but not to the degree for full localization to occur. This agrees with earlier results by
Brooks {\it et al.},\cite{Brooks_UNCO}
where the electronic structure of UO was calculated assuming itinerant $f$-electrons,
and which resulted in a calculated lattice parameter of 4.88 {\AA}, i.e. only slighty overbinding with respect
to the "experimental" value of 4.92 {\AA}. The SIC-LSD calculated lattice parameters of the $f^1$ and $f^2$
configurations, respectively 4.94 {\AA} and 4.99 {\AA}, indicate a slight overlocalization.
\begin{figure}
\begin{center}
\includegraphics[width=70mm,clip,angle=-90]{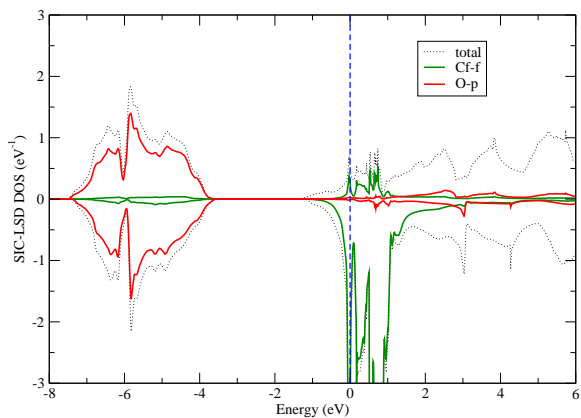}
\caption{
\label{DOSCFO}
Density of states for CfO in the ground state configuration with nine $f$-electrons localized on the Cf$^{3+}$ ion. The majority (minority) spin components
are displayed as positive (negative) values. The energy is measured in eV relative to the Fermi level, which falls amidst the minority Cf $f$ bands.
The localized Cf $f$ states are not shown.
}
\end{center}
\end{figure}
Overall, the calculated lattice parameters are in rather good agreement with the early experimental values.
However, as there is no convincing subsequent experimental evidence in support of the fact that monoxides
really exist in nature, one should not put too much weight on the agreement between theory and experiment.

For the monoxides beyond UO, the LSD configuration never becomes even
remotely energetically favourable. For NpO a tetravalent ground state (Np$^{4+}$$\equiv$ Np($f^3$)) is found,
whereas a trivalent ground state configuration is established for the remaining monoxides.
In their respective ground states, the monoxides are metallic, as can be seen from the non-zero density of states (DOS) at the Fermi energy
(column 4 of Table \ref{monoxidesTable}).
As a representative example, the DOS of CfO is depicted  in Fig. \ref{DOSCFO}.
The O atom has two unoccupied $p$-states,
and in the corresponding monoxide, the $p$-band can accomodate two electrons from the actinide atom through charge transfer and
hybridization, whilst the remaining
valence electrons, including the delocalized $f$-electrons, start filling the conduction band, with the Fermi level pinned
to the narrow $f$-band.
The ionic insulating picture would be realized if additional $f$-states preferred to localize, i.e. in
the divalent configuration.
In Table \ref{monoxidesTable} (column 3) the calculated energy differences, E$_{GS}$-E$_{II}$, between the ground
state configuration and the divalent
configuration are shown. It is clear that the nominal ionic A$^{2+}$O$^{2-}$scenario
does not become energetically favourable for any of the monoxides.
Nevertheless this energy difference is seen to decrease
from UO to AmO, and again from CmO to CfO, which is in agreement with the trends towards increasing localization due
to the actinide contraction. At AmO the trivalent configuration is thus only marginally more favourable than the divalent half-filled shell configuration, and
at EsO degeneracy occurs between the two configurations.
The overall non-ionic ground state picture for the monoxides that emerges from our SIC-LSD calculations
confirms the results from early molecular cluster calculations on the heavy actinide monoxides by
Gubanov {\it et al.},\cite{Gubanov}
where considerable covalency due to mixing of O-$p$ and A-$f$ orbitals was found.

\subsection{Actinide Sesquioxides}

\subsubsection{Background information}

Bulk phases of U$_2$O$_3$ as well as  Np$_2$O$_3$ do not exist in nature, and have never been synthesized, but thin films of Np$_2$O$_3$ have been found to
form on the surface of Np metal.\cite{Naegele} From Pu onwards, the sesquioxides are stable, and have been synthesized.
Beyond Pu, the sesquioxides crystallize in three different crystal structures, respectively the hexagonal La$_2$O$_3$ structure (A-form),
the cubic Mn$_2$O$_3$ structure (C-form), and the monoclinic Sm$_2$O$_3$ structure (B-form). Pu$_2$O$_3$ has been synthesized only in the
A- and C-forms.
The XPS measurements on sesquioxides from Pu$_2$O$_3$ to Cf$_2$O$_3$ have been reported,\cite{Courteix,Gouder_AmX,Veal} and the
absence of features at the Fermi level points towards the localized nature of the 5$f$ electrons in these compounds.
This indicates that they are semiconductors or insulators, in agreement with the ideal nominal picture of A$_{2}^{3+}$O$_{3}^{2-}$, although no values for
the energy gaps can be found in the literature.
\begin{table}
\caption{Actinide sesquioxide data: Column 2: Ground state configuration of actinide-ion. Column 3: Calculated energy gap, E$_{gap}$, (in eV).
Columns 4 and 5: Calculated, V$_{calc}$, and experimental\cite{Pearson}, V$_{exp}$, equilibrium volume(s) in in units of $\AA^3$ per formula unit. }
\begin{ruledtabular}
\begin{tabular}{|c|c|c|c|c|}
Compound  &  Ground state                 &   E$_{gap}$   &  V$_{calc}$  & V$_{exp}$     \\
\hline
\multicolumn{5}{|c|}{C-type sesquioxides} \\
\hline
\hline
  U$_2$O$_3$ & f$^2$ (U$^{4+})$    & 0.00  & 83.17  & - \\
\hline
  Np$_2$O$_3$ & f$^3$ (Np$^{4+})$  & 0.00  & 84.40  & - \\
\hline
  Pu$_2$O$_3$ & f$^5$ (Pu$^{3+})$  & 0.75  & 89.42  & 82.73   \\
\hline
  Am$_2$O$_3$ & f$^6$ (Am$^{3+})$  & 0.44  & 88.54  & 83.64 \\
\hline
  Cm$_2$O$_3$ & f$^7$ (Cm$^{3+})$  & 0.32  & 86.98  & 83.10  \\
\hline
  Bk$_2$O$_3$ & f$^8$ (Bk$^{3+})$  & 0.38  & 83.41  & 80.63 \\
\hline
  Cf$_2$O$_3$ & f$^9$ (Cf$^{3+})$  & 0.47  & 82.60  & 79.59  \\
\hline
\multicolumn{5}{|c|}{A-type sesquioxides } \\
\hline
\hline
  Pu$_2$O$_3$ & f$^5$ (Pu$^{3+})$  & 2.43 & 74.06  & 75.49   \\
\hline
  Am$_2$O$_3$ & f$^6$ (Am$^{3+})$  & 2.54 & 73.34  & 74.73 \\
\hline
  Cm$_2$O$_3$ & f$^7$ (Cm$^{3+})$  & 3.07 & 72.40  & 74.53 \\
\hline
  Bk$_2$O$_3$ & f$^8$ (Bk$^{3+})$  & 2.73 & 70.10  & 72.71 \\
\hline
  Cf$_2$O$_3$ & f$^9$ (Cf$^{3+})$  & 1.78 & 69.33  & 71.43 \\
\end{tabular}
\end{ruledtabular}
\label{sesquioxideTable}
\end{table}

There exist relatively few calculations of the electronic structure of the actinide sesquioxides.
Prodan {\it et al.}\cite{Prodan_Pu2O3} have studied Pu$_2$O$_3$ (A-form) with the help of hybrid density functional theory, comparing a number of functionals. They have found that unlike in the LSD and GGA approximations,
using the Heyd-Scuseria-Ernzerhof (HSE) screened Coulomb functional leads to an insulating anti-ferromagnetic
solution, in good agreement with experiment, with the calculated gap value of 2.78 eV (3.50 eV for the
Perdew-Burke-Ernzerhof (PBE0) hybrid functional).
The antiferromagnetic insulating nature of Pu$_2$O$_3$ has similarly been retrieved from electronic structure calculations
 by respectively Jomard  {\it et al.}\cite{Jomard} and Sun {\it et al.}\cite{Sun}, based on the (LDA/GGA)+U approximation.
Their gap values are found to depend strongly on the value chosen for the electron-electron interaction U.
For U=4 eV, and depending on the details of the functional, energy gaps ranging from 1 to 2 eV were obtained.

\subsubsection{SIC-LSD electronic structure}

\begin{figure}
\begin{center}
\includegraphics[width=90mm,clip]{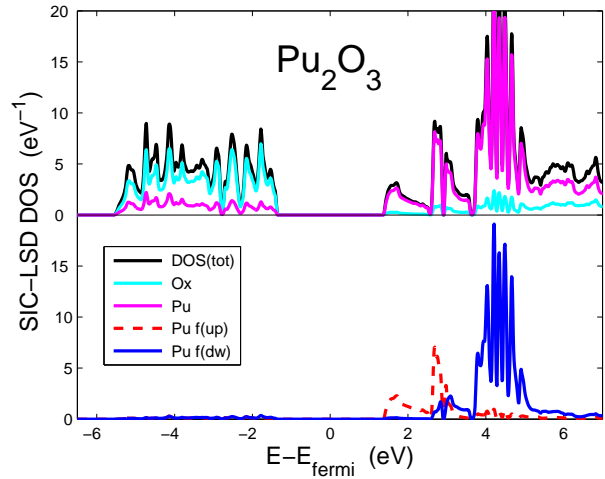}
\caption{
\label{Pu2O3}
Density of states for A-type Pu$_2$O$_3$ in the trivalent ground state configuration.
The energy is in eV relative to the Fermi level, which is situated midgap. The DOS is in units of eV$^{-1}$ per formula unit.
Upper panel: Total (black), Pu (violet), and O (light blue) DOS's. Lower panel:
The corresponding Pu-$f$ majority (red), and minority (blue) spin decomposed DOS's. Only the itinerant $f$-states are shown.
}
\end{center}
\end{figure}

In the present work the electronic structures of the sesquioxides from U$_2$O$_3$ to Cf$_2$O$_3$ has been calculated
for both the cubic C-form and the hexagonal A-form. The results are summarized in Table \ref{sesquioxideTable}.
The hexagonal A-type structure has space group P\={3}m1 (no. 164). The unit cell contains one formula unit with one
Oxygen at the origin, two Oxygens at $\pm$(1/3, 2/3, z$_O$) and two actinide atoms at $\pm$(1/3, 2/3, z$_A$).
While the c/a ratio has been measured for all the actindes from Pu to Cf, the internal parameters are only known for
Pu (z$_O$ = 0.6451 and z$_{Pu}$ = 0.2408).\cite{Pearson} Hence we have performed our calculations for this structure,
with the experimental c/a ratios but using the internal parameters of Pu$_2$O$_3$ for all the other actinide
sesquioxides.\cite{Wulff} The C-type structure, also known as bixbyite, has been approximated by the fluorite
AO$_2$ structure, with 1/4 of
the O atoms removed, i.e. from four formula units in a conventional simple cubic supercell the oxygen atoms at the
origin and the cube center were replaced by empty spheres.
No relaxation of the atomic coordinates has been attempted.
The trivalent actinide configuration is found to be energetically favourable for all the compounds, apart from
U$_2$O$_3$ and Np$_2$O$_3$, for which   the tetravalent ground state configuration is preferred. Incidentally, these are also the only
sesquioxides that do not occur in nature.
In the trivalent ground state, the sesquioxides are found to be insulators, with energy gaps of around 0.5 eV for the C-form, and around
2.7 eV for the A-form. The corresponding DOS for the A-type Pu$_2$O$_3$ is shown in Fig. \ref{Pu2O3}.

Only the ferromagnetic ordering was investigated for the cubic structure, whilst only the antiferromagnetic ordering
of spins was studied for the hexagonal structures,
and the values for the insulating gaps in Table
\ref{sesquioxideTable} refer to these respective magnetic orderings.
The bulk moduli for A-type Pu$_2$O$_3$, Am$_2$O$_3$, Cm$_2$O$_3$, Bk$_2$O$_3$, and Cf$_2$O$_3$ are similar in magnitude, calculated
to be respectively, 158, 158, 168, 166, and 174 GPa.
No experimental measurements of the bulk moduli seem to exist, but the value for Pu$_2$O$_3$ lies within the range of values
obtained by Prodan {\it et al.}\cite{Prodan_Pu2O3} and Jomard  {\it et al.}\cite{Jomard} for a suite  of different functionals (from 110 GPa to 181 GPa).
As can be seen from Table \ref{sesquioxideTable}, the calculated equilibrium volumes of the A-type sesquioxides are
found to be in good agreement with experiment. This agreement is somewhat less satisfying for the C-type sesquioxides,
where the volume is overstimated by up to 4 \%, which
might be related to our approximate treatment of the actual bixbyite structure.

\subsection{Actinide Dioxides}

\subsubsection{Background information}

The actinide dioxides from U to Cf have all been synthesized, crystallizing in the fluorite (CaF$_2$) structure. They are also the
most relevant systems for applications and, as a consequence, are the most studied actinide oxide compounds, both experimentally and theoretically.
\cite{Boettger_PuO2,Jomard,Boettger_UO2,Yun,Sun,Dudarev,Yin,Petit_PuO2,Prodan_AcOxide,Kelly_ionicity,Kelly_AcOx,Johansson_AcOxides,Troc,Veal,Baer_UO2,Courteix,Fournier}
On the experimental side, a comprehensive summary of their electronic, magnetic, transport and optical properties can be
found in Tro\'{c} {\it et al.}.\cite{Troc}
The absence of features at the Fermi level in the observed XPS spectra\cite{Veal} indicates that all the dioxides are semiconductors or insulators.
However, specific  information regarding the gap is only known for UO$_2$, NpO$_2$, and PuO$_2$. A combined XPS and BIS spectrum for UO$_2$ finds an O(2p) $\rightarrow$ U(6d) gap
of about 5 eV.\cite{Baer_UO2} This value for the energy gap is similar to the one observed for ThO$_2$, but with the difference that in UO$_2$
two occupied rather well localized $f$-states are situated in the gap, with a 5$f^2$ $\rightarrow$  5$f^1$6$sd$ transition energy of 2-3 eV.\cite{Baer_UO2,SchoenesFar}
As the nuclear charge increases from Th to Pu, the $f$-states move to lower energies. In ThO$_2$ the empty $f$-states are situated in the Th-$sd$ derived conduction
band. In UO$_2$ and NpO$_2$ the $f$-states are occupied and situated in the energy band gap, whereas in PuO$_2$
they are situated at the top of the O-$p$ derived valence band.\cite{Courteix}
The measured activation energies, E$_a$=0.2 eV in UO$_2$, E$_a$=0.4 eV in NpO$_2$, and E$_a$=1.8 eV in PuO$_2$, confirm this trend.\cite{Fournier}

Electronic structure calculations, in particular for UO$_2$ and PuO$_2$, have shown that the itinerant $f$-electron picture is not adquate for describing these compounds. It has emerged that the LSD approximation wrongly predicts metallic
behaviour,\cite{Boettger_PuO2,Jomard,Boettger_UO2} demonstrating the need to go beyond the homogeneous electron gas in describing the strong on-site $f$-electron correlations. So far, a number of calculations have shown
that an insulating solution can be obtained when using improved descriptions of electron correlations such as those contained in
the LDA(GGA)+U\cite{Yun,Sun,Jomard,Dudarev},
DMFT\cite{Yin}, SIC-LSD\cite{Petit_PuO2} and hybrid functional methodologies.\cite{Prodan_AcOxide}

\subsubsection{SIC-LSD electronic structure}

In the present paper the electronic structures of the dioxides from UO$_2$ to CfO$_2$ has been calculated using the SIC-LSD
method.
As can be seen from Table \ref{dioxidesTable}, a tetravalent ground state configuration is found
for all the dioxides, except UO$_2$ where the pentavalent (U($f^1$)) configuration is energetically more favourable.
Actually, the pentavalent U($f^1$) and tetravalent U($f^2$) configurations are energetically close, with the $f^1$ configuration
being more favourable by some 100 meV.
Our calculations refer to T=0 K, whereas the experimental evidence, which clearly indicates a tetravalent (insulating) UO$_2$, mostly refers
to room temperature conditions.
Experiments also indicate a lattice expansion with temperature,\cite{Ogden,Kang} but the $f^1$ to $f^2$ localization transition
that our calculations seem to predict has not been observed experimentally.
The prediction of an U-$f^1$ goundstate configuration could possibly be related to the
tetrad effect (multiplet formation energy),\cite{Jorgensen,Nugent,Xia} which is ignored in our calculation, and which favours
the $f^2$ configuration over the $f^1$ configuration.
In the following, when comparing to experiment, we will be referring
to the tetravalent UO$_2$ configuration as the ground state configuration.

All the dioxides (including the tetravalent
UO$_2$) are predicted to be insulators. The AFII magnetic ordering ({\it i.e.} ferromagnetically ordered planes stacked
anti-ferrormagnetically along the $[111]$  crystal
direction) has been assumed in the dioxide calculations to which the SIC-LSD data in Table \ref{dioxidesTable} refer.
With respect to the band gap of UO$_2$, a value of E$_{gap}$=2.6 eV is found, which is smaller than the experimentally
observed $p$$\rightarrow$$d$ gap, and which should not be compared to the optical gap referred to in the 
GGA+U (1.8 eV), and HSE (2.4 eV) calculations.
The SIC-LSD, being a one-electron ground state theory, does not give accurate
removal energies of localized states, on account of the missing screening and relaxation effects, predicting them at
too high binding energies, in disagreement with spectroscopy.
As a result, the SIC-LSD  calulations do not reproduce the correct position of the occupied $f$-peak
which is expected to be situated in the gap for
both UO$_2$ or NpO$_2$. The Mott-insulating character ($f$$\rightarrow$$f$ transition) of the early dioxides\cite{Prodan_AcOxide,Yun}
is thus not reproduced in our calculations, which instead describe the entire series as charge transfer insulators ($p$$\rightarrow$$f$ transition).
From PuO$_2$ onwards, the occupied $f$-states are situated at or below the valence band maximum, and the charge transfer picture becomes
adequate for describing the nature of the gap. There exist only a few calculations dealing with the electronic structure
of the heavy actinide dioxides.\cite{Gubanov,Prodan_AcOxide,Kelly_ionicity} In the following we make a detailed comparison with the
recent results obtained by Prodan {\it et al.}\cite{Prodan_AcOxide} using the hybrid functional theory.
\begin{table}
\caption{Actinide dioxide data: Column 2: ground state configuration. Columns 3 and 4: Energy gaps, E$_{gap}^{SIC}$ (this work) and E$_{gap}^{HSE}$ (reference \onlinecite{Prodan_AcOxide}), respectively (in units of eV).
Columns 5 and 6: calculated, a$_0^{SIC}$, and experimental a$_0^{exp}$ lattice parameters (in {\AA}). Column 7 and 8: calculated, B$_0^{SIC}$, and experimental, B$_0^{SIC}$, bulk modulii (in units of GPa).}
\label{dioxidesTable}
\begin{ruledtabular}
\begin{tabular}{|c|c|c|c|c|c|c|c|}
   & Config. & E$_{gap}^{SIC}$ & E$_{gap}^{HSE}$  & a$_0^{SIC}$ & a$_0^{{exp}^a}$ & B$_0^{SIC}$ & B$_0^{exp}$  \\
\hline
  UO$_2$ & f$^1$ (U$^{5+}$)                &  0    & 2.4 & 5.40       & 5.470                & 219& 207$^b$ \\
\hline
  UO$_2$ & f$^2$ (U$^{4+}$)                &  2.6  & 2.4 & 5.47       & 5.470                & 219& 207$^b$ \\
\hline
  NpO$_2$ & f$^3$ (Np$^{4+})$              &  2.3  & 3.1 & 5.46       & 5.433                & 217 & 200$^b$  \\
\hline
  PuO$_2$ & f$^4$ (Pu$^{4+})$              &  1.2  & 2.7 & 5.44       & 5.396                & 214 & 178$^b$ \\
\hline
  AmO$_2$ & f$^5$ (Am$^{4+})$              &  0.8  & 1.6 & 5.42       & 5.374                & 209 & 205$^b$  \\
\hline
  CmO$_2$ & f$^6$ (Cm$^{4+})$              &  0.4  & 0.4 & 5.37       & 5.359                & 212 & 218$^c$  \\
\hline
  BkO$_2$ & f$^7$ (Bk$^{4+})$              &  1.0  & 2.5 & 5.36       & 5.331                & 221 & - \\
\hline
  CfO$_2$ & f$^8$ (Cf$^{4+})$              &  0.6  & 2.0 & 5.36       & 5.310                & 210 & - \\
\hline
\end{tabular}
\end{ruledtabular}
$^a$Reference \onlinecite{Pearson}
$^b$Reference \onlinecite{Idiri}
$^c$Reference \onlinecite{Dancausse}
\end{table}

\begin{figure}
\begin{center}
\includegraphics[width=80mm,clip,angle=0]{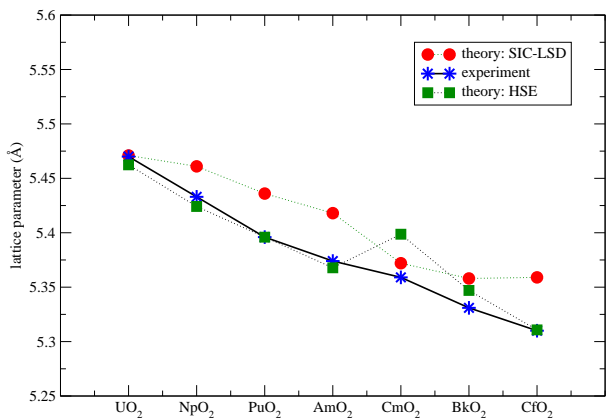}
\caption{
\label{alatAcO2}
Lattice parameters of the actinide dioxides. Experimental values\cite{Pearson} (blue stars) are compared to the theoretical SIC-LSD values (red circles), and
the theoretical HSE values (green squares) read from the corresponding figure 1 in reference \onlinecite{Prodan_AcOxide}.
}
\end{center}
\end{figure}

Both the SIC-LSD and the hybrid functional calculations determine the antiferromagnetic ground state configuration
as energetically most favourable.  As can be seen in Table \ref{dioxidesTable}, our calculated
energy gaps are smaller than the corresponding HSE gaps, but the trends agree, i.e. the gaps decrease from PuO$_2$ to
CmO$_2$, and then again from BkO$_2$ onwards. This reflects the gradual progression towards lower energy of the unoccupied $f$-states, with increasing atomic number.
The calculated lattice parameters are in very good agreement with experiment, within  1 \%, as seen in both
Table \ref{dioxidesTable} and
figure \ref{alatAcO2}. We should note here that the experimentally observed lattice parameters refer to room temperature measurements,\cite{Pearson} and are thus
on average larger than the values one would expect for T=0 K by approximately 0.011 \AA.\cite{Benedict}
The consistent overestimation of the lattice parameters is quite common for SIC-LSD, which has tendency to slightly overlocalize.

The overall experimental situation with regard to CmO$_2$ is still not fully understood.\cite{Soderholm}
It has been shown that, for a range of off-stoichiometric compounds CmO$_{2-x}$, both the lattice parameter and
magnetic moment increase with $x$.\cite{Nave,Peterson,Noe} Given the expected moments of Cm$^{3+}$ (5$f^7$; $\mu_{eff}$=7.94 $\mu_B$) and Cm$^{4+}$ (5$f^6$; $\mu_{eff}$=0 $\mu_B$)
it was noted that this trend
indicates an increase in the number of Cm$^{3+}$ impurities as we move away from stoichiometry.
Based on the ionic picture, one accordingly expects the Cm$^{4+}$ ground state configuration for the stoichiometric compound,
which however seems to be contradicted by the fact that in samples very close to stoichimetry,\cite{Morss} susceptibility
measurements give the effective moment as high as $\mu_{eff}$= 3.36 $\mu_B$.
Attempts to explain this discrepancy, range from a possible existence of an impurity phase, not registered in the
experiments, to assuming a covalent, rather than an ionic, picture for the ground state.\cite{Soderholm}
It has also been conjectured that a more complex ground state which includes
some admixture of the excited $J=1$ state of Cm$^{4+}$ might be required.\cite{Morss}

The lattice parameter measurements yield two different values for CmO$_2$,
depending on whether the short-lived curium isotope of mass number 244 is used ($a=5.372$ {\AA}),\cite{Asprey} or the
long-lived curium isotope of mass number 248 is used ($a=5.359$ {\AA}).\cite{Peterson} The reason for this discrepancy has been traced
to the self-radiation induced expansion that occurs with the considerably more active $^{244}$Cm isotope.\cite{Peterson,Noe}
It is the $^{248}$CmO$_2$ lattice parameter, i.e. without the effect of self-heating, that should be of relevance when comparing to the calculated lattice parameters.

\begin{figure}
\begin{center}
\includegraphics[width=90mm,clip,angle=0]{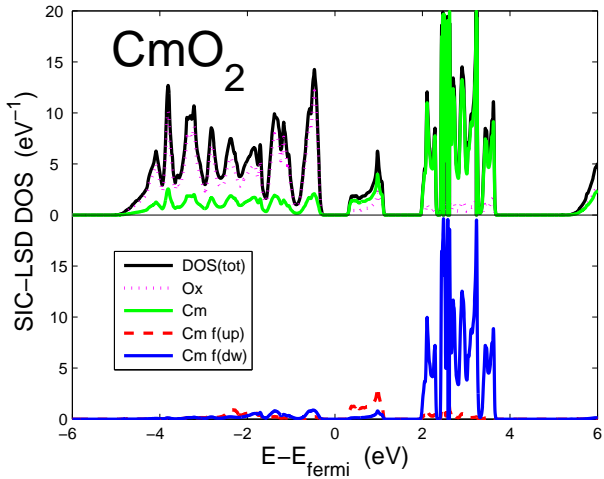}
\caption{
\label{CmO2}
DOS of CmO$_2$.
Upper panel: Total (black), Cm (green), and O (red dotted) DOS's. Lower panel:
The corresponding Cm-$f$ majority (red), and minority (blue) spin decomposed DOS's. Only the itinerant $f$-states are shown.
}
\end{center}
\end{figure}

From the SIC-LSD calculations we find the ground state of CmO$_2$ to be tetravalent, and the corresponding lattice parameter is in good agreement with experiment.
In figure \ref{CmO2}  we show the DOS of CmO$_2$ in the Cm$^{4+}$ ($f^{6}$) ground state configuration. The compound is found to be insulating, with the
Fermi level positioned between the completely filled O-$p$ band, and the one remaining delocalized empty majority $f$-spin state that
strongly hybridizes with the O-$p$ states. The resulting electronic structure is thus quite different from the hybrid functional picture,
where the HSE applied by Prodan et al. results in metallic CmO$_2$. The authors suggest a covalent picture with significant Cm$^{3+}$ character,
as a result of the Cm trying to achieve the stable half-filled $f$-shell configuration. Consequently some of the O $p$-states are charge transferred
to the Cm-ion, resulting in the Fermi level cutting accross the top of the hybridized O $p$ - Cm $f$ band.
However, unlike for all the other actinide dioxides, the lattice parameter of CmO$_2$, calculated using the HSE functional,
deviates considerably from the measured value, as can be seen from Fig. \ref{alatAcO2}.

With respect to the SIC-LSD calculations, one could envisage that a Cm$^{3+}$ configuration in CmO$_2$ could be
energetically favourable, given the associated half-filled $f$-shell. Due to one less $f$-state
taking part in bonding, as compared to the Cm$^{4+}$, this in return
could explain the slight increase in the experimentally observed lattice parameter of CmO$_2$.
As it turns out, however, the total energy 
of the trivalent configuration is higher by 0.68 eV, indicating that the Cm$^{4+}$ configuration is energetically very
stable. The reason for this stability is related to the fact that localizing an additional electron in CmO$_2$ results
in the Fermi level moving down into the $p$-band.
This implies a depopulation of the $p$-band, as charge is transferred to the low lying $f$-levels, and the
associated loss in respectively the Madelung and hybridization energies is significantly
larger than the gain in the $f$-localization energy. This charge transfer picture is similar to the picture found in
the hybrid functional theory.
However, in variance to the conclusion of that work, we find the corresponding trivalent configuration to be
energetically unfavourable.

Although our calculated ground state of CmO$_2$ finds a localized $f^6$ Cm
ion, the system turns out to be magnetic.
The ideal $f^6$ ion has $J=0$ and hence also
 $<S_z> =0$ and $<L_z> =0$, but this state is a linear combination of
several 6-electron Slater determinants, which we cannot treat in our SIC-LSD
scheme.
Rather, it may be represented by a single Slater determinant having
 $<S_z> =3$ and $<L_z> =-3$ (simulating antiparallel $S=3$ and $L=3$), which
constitutes the starting point for our calculations before iteration to
self-consistency.
Spin-orbit coupling and hybridization significantly distort this initial
configuration, in fact almost quenching the orbital moment.
The total spin and orbital moment projections along the $z$-axis are calculated to be
$<S_z> = 2.73$ and $<L_z> =-0.25$
(in units of $\hbar$). This comes from contributions from the localized
$f^6$ ion
( $<S_z> = 2.76$ and $<L_z> =-1.15$) and from the delocalized f-band states
( $<S_z> =-0.03$ and $<L_z> =+0.93$). The latter are largely tails of the O
p-band states, which inside the atomic sphere around Cm attain
$f$-character.
The total number of delocalized $f$-character states is 0.86, leading to a total
$f$-count of 6.86, which reflects a significant $f$-hybridization with the O
p bands.
Assuming now, that the magnetic moment is given as $\mu=<2S_z+L_z>\mu_B$, we
arrive at a total moment of $\mu =5.21 \mu_B$ per Cm atom, which exceeds the
experimental
moment of $\mu_{eff}=3.36 \mu_B$,\cite{Morss} however demonstrating that
the insulating state of CmO$_2$ may be magnetic.
The quoted experimental moment is extracted from the large temperature
magnetic susceptibility, while the present theory is valid only at $T=0$.
Furthermore, as stated, the fact that quantum fluctuations (more than one
Slater
determinant representing the localized  $f^6$ shell), are not possible within
the SIC-LSD approach, might be a serious limitation with respect to a proper
description of magnetic properties. Thermal fluctuations can be considered
using the so-called local SIC (LSIC) approach, implemented in the multiple
scattering theory, in combination with the coherent potential approximation
(CPA) and disordered local moment (DLM) theory.\cite{Lueders_LSIC,Hughes}

\subsubsection{Oxidation and Reduction Energies}

Uranium metal readily oxidizes to form a range of super-stoichiometric oxide compounds, UO$_{2+x}$, as well as
stoichiometric compounds, UO$_2$, U$_3$O$_8$, UO$_3$, and U$_2$O$_5$.
Sub-stoichiometric UO$_{2-x}$ exists, but U$_2$O$_3$ for example is not found in the literature.
In this section we will investigate oxidation/reduction of the actinide dioxides, based on the SIC-LSD total energies involved in
the different delocalization/localization transitions of the $f$-electrons.
To model the oxidation process
\begin{eqnarray}
\mbox{A}\mbox{O}_2+\mbox{$\frac{1}{8}$O}_2 &\rightarrow & \mbox{A}\mbox{O}_{2.25} ,
\label{ox}
\end{eqnarray}
one additional O is introduced into a four formula units AO$_2$ supercell, i.e. we define
the oxidation energy
\begin{eqnarray}
E^{ox}=\frac{1}{4}\left [ E(\mbox{A}_4\mbox{O}_9)-
E(\mbox{A}_4\mbox{O}_8)-\frac{1}{2}E(\mbox{O}_2)\right ].
\label{Eox}
\end{eqnarray}
The CaF$_2$ structure is assummed to remain undistorted, i.e.
relaxation effects are  not taken into account. For the AO$_2$ reduction process
we consider the reaction
\begin{eqnarray}
\mbox{A}\mbox{O}_2 &\rightarrow & \mbox{A}\mbox{O}_{1.5}+\mbox{$\frac{1}{4}$O}_2,
\label{red}
\end{eqnarray}
In the supercell used for modelling the reaction, two O atoms are removed from the four formula units AO$_2$ supercell. Correspondingly,
the reduction energy is
\begin{eqnarray}
E^{red}=\frac{1}{4}\left [E(\mbox{A}_4\mbox{O}_6)+E(\mbox{O}_2)-
E(\mbox{A}_4\mbox{O}_8)\right ].
\label{Ered}
\end{eqnarray}
The compound A$_4$O$_6$ is basically a two formula unit
supercell of the sesquioxide A$_2$O$_3$ in the previously described approximation to the bixbyite structure.

The supercell total energies E(A$_4$O$_8$) and E(A$_4$O$_6$) have been evaluated at the ground state valency configurations
established in the preceding sections for respectively AO$_2$ and A$_2$O$_3$.
For A$_4$O$_9$, the SIC-LSD calculations find the following ground state configurations: U$_4^{5+}$O$_9$, Np$_4^{5+}$O$_9$, Pu$_2^{4+}$Pu$_2^{5+}$O$_9$,
Am$_4^{4+}$O$_9$, Cm$_3^{4+}$Cm$^{5+}$O$_9$/Cm$_4^{4+}$O$_9$, Bk$_4^{4+}$O$_9$, and Cf$_4^{4+}$O$_9$. The corresponding total energy minima
E(A$_4$O$_9$) are used in the oxidation energy calculations of Eq.(\ref{Eox}). Compared to the tetravalent dioxide, A$_4$O$_8$, we observe that
in A$_4$O$_9$, the valency trend is from pentavalent in the early actinides to tetravalent in the late actinides, indicating
that the inclusion of O impurities into the dioxide leads to $f$-electron delocalization in the early actinide oxides U$_4$O$_9$, Np$_4$O$_9$,
and Pu$_4$O$_9$, whereas from Am$_4$O$_9$ onwards the compounds remain tetravalent. On the other hand the fact that all the sesquioxides, A$_4$O$_6$, prefer the
trivalent ground state configuration shows that the removal of O from the dioxide leads to localization of an additional
$f$-electron.
Whether the oxidation (Eq.(\ref{ox})) or reduction (Eq.(\ref{red}))  process actually takes place depends on the thermodynamic conditions,
in particular the O chemical potential.

\begin{figure}
\begin{center}
\includegraphics[width=70mm,clip,angle=0]{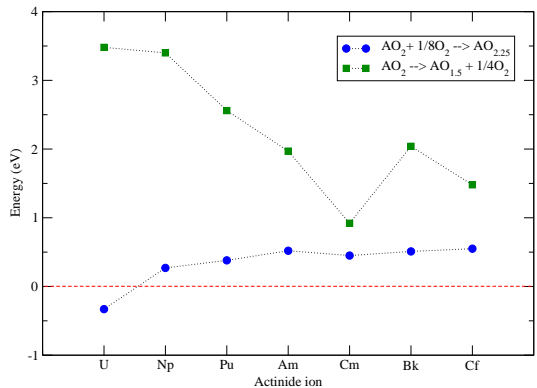}
\caption{
\label{redox}
Oxidation (circles) and reduction (squares) energies of the actinide dioxides according to the definitions in Eqs. (\ref{Eox}) and (\ref{Ered}).
Energies are in eV per AO$_2$ unit cell.
}
\end{center}
\end{figure}

The results for the calculated oxidation and reduction energies are presented in Fig. \ref{redox}. A binding energy of 7.2 eV has been adopted for
the O$_2$ molecule (relative to the non spin-polarized O atoms. The atomic spin-polarization energy\cite{Peltzer} is 1.14 eV
per O atom). The accuracy of the
energy differences is not expected to be very high due to the ASA and the neglect of structural optimizations. Note,
however, that for Pu$_2$O$_3$, our calculated
value of 2.6 eV/PuO$_2$ for the reduction process, Eq.(\ref{Ered}), lies within the range of LDA+U/GGA+U ($U = 4$ eV) values
calculated by Jomard {\it et al.}\cite{Jomard}, 2.59/2.05 eV/PuO$_2$,
and Sun {\it et al.}\cite{Sun}, 1.50/2.00 eV/PuO$_2$ (data points taken from their figure).
Nevertheless, here we will only concentrate on trends, which 
come out quite clearly.

It emerges that $E^{ox}$ increases slightly and $E^{red}$ decreases as we move through the actinide series, which reflects the increasing f-electron
binding energy. In Uranium an f-electron is readily promoted into the valence band to facilitate uptake of super-stoichiometric oxygen, while this becomes increasingly
more difficult with heavier actinides. On the other hand the U
f-manifold is very reluctant to take on an extra f-electron to form the 3+ ion on reduction, while this becomes easier for the later actinides. The jump in reduction
energy at Bk reflects the relative stability of the half-filled $f^7$ shell in BkO$_2$.

According to the calculated energy differences, the only compound that oxidizes exothermically is UO$_2$, whilst for all other actinide dioxides, the energy
balance is positive and increasing with actinide number. This result is in overall good agreement with experimental data.
UO$_{2+x}$ has been synthesized up to x=0.25 (U$_4$O$_9$). The crystal structure remains fluorite based, with the excess O atoms situated
at interstitial sites.\cite{Allen_uo2} With respect to Np, the only known stable binary oxide, apart from NpO$_2$, is Np$_2$O$_5$, with a structure
that can no longer be described as underlying fluorite.\cite{Forbes}
In particular, the fluorite derived Np$_4$O$_9$ does not seem to exist. With respect to PuO$_{2+x}$, the present
calculations confirm the stability of PuO$_2$, in line with the majority of experimental investigations
on this material. The same conclusion was reached in an earlier SIC-LSD study\cite{Petit_PuO2} where the possible discovery of a higher composition binary oxide,
PuO$_{2+x}$ (x $\leq$ 0.27)\cite{Haschke} was discussed. Beyond Pu there have been no reports of super-stoichiometric oxides.
The positive slightly increasing oxidation energies in Fig. \ref{redox} confirm that the dioxide for all the
trans-plutonium actinides is increasingly stable against oxidation.

Concerning the reduction reaction in Eq.(\ref{red}), it is  found that the energy balance always favors the dioxide.
With respect to the early actinides U and Np the large value of the reduction energy is in agreement
with the experimental fact that neither U$_2$O$_3$ nor Np$_2$O$_3$ exists. The sesquioxides from Pu$_2$O$_3$ to Cf$_2$O$_3$ do exist,
however they are according to
our calculations in principle unstable in air towards further uptake of oxygen and formation of the dioxide.
In practice there may be appreciable barriers to the actual transformation,\cite{Baybarz} which furthermore may be
influenced by thermodynamic conditions (temperature and pressure)
not considered here.

\section{Summarizing discussion}

From the study of the stable binary actinide oxides, a clear picture emerges that links the degree of oxidation to the
degree of $f$-electron localization. In the early actinides, the $f$-electrons are less bound to the actinide ions which
translates into valencies as high as 5+ and 6+ for U-oxides for example. As one progresses through the actinide series,
the $f$-electrons
become increasingly tightly bound to the actinide ion, and eventually for Cf only the 3+ valency occurs naturally.
The actinide ions play an
active role in accomodating extra O, as their localized $f$-electrons can act as electron reservoirs for the highly electronegative O-ion.
In other words, whether the oxidation is favoured depends on the willingness of the actinide to delocalize further $f$-electrons.
From the SIC-LSD study we find that the divalent configuration is never favoured, except maybe for EsO. Consequently, the monoxide never forms
as the third electron is readily delocalized and made available for oxidation, which leads to the formation of trivalent sesquioxides. Whether
the oxidation progresses further to produce the corresponding dioxide depends on delocalization of the 4th electron on the actinide ion. As we have seen
this delocalization is less likely to happen for the late actinides, where the $f$-electrons tend to be more localized, and where experiment shows
that the synthesis of CfO$_2$ takes place under powerful oxidation conditions, e.g. in the presence of high pressure
molecular or atomic oxygen.\cite{Baybarz} For the early actinides, on the other hand, the oxidation to the dioxide
occurs readily,
and for example U$_2$O$_3$ and
Np$_2$O$_3$ do not exist naturally. Finally, the further oxidation from dioxide to higher oxide only occurs for UO$_2$, where both the
5+ (U$_2$O$_5$) and 6+ (UO$_3$) U-valencies exist, and for NpO$_2$ where the 5+ valency is obtained in Np$_2$O$_5$. 
We can speculate here as to whether the fact that we actually determine a marginally pentavalent groundstate for UO$_2$ (rather than the
well established tetravalent configuration) is an indication that
an additional electron $f$-electron is on the brink of delocalization, thus favouring the formation of higher U-oxides. The remaining dioxides are stable
with respect to oxidation, as the gain in delocalizing the 5th electron is no longer favorable compared to the corresponding loss in SIC energy.
\begin{table}
\caption{
Groundstate valency configurations of the actinide and lanthanide oxides. Bold lettering indicates that the corresponding compound actually exists in nature.}
\label{review}
\begin{ruledtabular}
\begin{tabular}{|c|c|c|c|c|c|c|c|c|}
A  & U & Np & Pu & Am & Cm & Bk & Cf&Es \\
\hline
AO & 5+ & 4+ & 3+ & 3+ & 3+ & 3+ & 3+ & { \bf 3+/2+}  \\
\hline
A$_2$O$_3$ & 4+ & 4+ & { \bf 3+} & { \bf 3+} & {\bf 3+} & { \bf 3+} & { \bf 3+}  & - \\
\hline
AO$_2$ & {\bf 5+/4+} & {\bf 4+} & {\bf 4+} & {\bf 4+} & { \bf 4+} & {\bf 4+} & {\bf 4+} &-  \\
\hline
\hline
R  & Nd & Pm & Sm & Eu & Gd & Tb & Dy & Ho \\
\hline
RO & 3+ & 3+ & {\bf 3+/2+}  & {\bf 2+} & 3+ & 3+ & 3+ & 3+  \\
\hline
R$_2$O$_3$ & {\bf 3+} & { \bf 3+} & {\bf 3+} & {\bf 3+} & {\bf 3+} & {\bf 3+} & {\bf 3+} & {\bf 3+} \\
\hline
RO$_2$ & 4+ & 4+ & 3+  & 3+ & 3+ & {\bf 4+} & 3+  & 3+ \\
\end{tabular}
\end{ruledtabular}
\end{table}

In Table \ref{review} we have collected our results of the SIC-LSD total energy calculations for the actinide oxides
(the upper 4 rows). The numbers indicate the ground state
valency configurations that we have determined for a given compound. Bold large letters are used to indicate those compounds
that actually do exist in nature. It clearly emerges that only in those cases where the calculated ground state valency
agrees with the nominal charge expected from an ideal ionic picture, does the corresponding oxide seem to form. In other
words, those oxides where our calculations predict a valency configuration that is not in agreement with simple
charge counting will not form naturally, and the excess/shortage of charge indicates that oxidized/reduced compound
will be favoured instead. These trends emphasize the very ionic nature
of bonding in the actinide oxides.

It is interesting to compare these trends in oxidation of the actinides to the corresponding behaviour of the lanthanides.\cite{Petit_reoxide_PRB,Petit_reoxide_Springer}
 In the latter the 4$f$ electrons are
overall more tightly bound to the lanthanide ion, which will favour lower oxidation numbers, compared to the spatially more extended 5$f$ electrons.
The corresponding calculated valency configurations\cite{Petit_reoxide_PRB} are shown in Table \ref{review} (the lower 4 rows).
The effect of the increased localization can be observed from the fact that a number of the lantanide monoxides actually exist, and
it emerges from SIC-LSD studies that especially for the half-filled and filled shell $f$-electron systems, i.e. for EuO and YbO,
the divalent configuration is energetically most favorable. All the lanthanide sesquioxides occur in nature, and the ground state configuration
of the corresponding lanthanide ions is trivalent. Even though the 4$f$-electrons tend to be more localized, the 3+ valency occurs naturally,
which is in agreement with the trivalent configuration being the most favourable for the lanthanide metals. With respect to further oxidation to the
tetravalent dioxide, only CeO$_2$, PrO$_2$ (not shown in Table \ref{review}), and TbO$_2$, are found to occur naturally or (for TbO$_2$)
under high O pressure, indicating that an additional $f$-electron
only delocalizes in the very early lanthanides, i.e. in Ce and Pr, where the $f$-electrons are less tightly bound, or in the middle of the series,
where tetravalent TbO$_2$ is favoured by the half-filled shell configuration.\cite{Petit_reoxide_PRB} Higher oxidation
numbers than IV do not exist for the lanthanide oxides. Again we observe that agreement between nominal charge and calculated
ground state valency is
required for a given compound to exist.


\section{Conclusion}

We have studied the electronic structure of actinide oxides and specifically monoxides, sesquioxides and dioxides, within
the ab initio SIC-LSD band structure method. By studying the oxidation and reduction reactions we have been able to conclude
that the dioxides, from Np onwards, are the most stable compounds among the studied actinide oxides.
Our study reveals a strong link between the preferred oxidation number
and the degree of localization which is confirmed by comparing to the ground state configurations of the corresponding
lanthanide oxides. The ionic nature of the actinide oxides is reflected in that only those compounds can form where
the calculated ground state valency agrees with the nominal
valency expected from a simple charge counting.

\section * {Acknowledgment}
This research used resources of the Danish Center for Scientific Computing (DCSC) and of the National Energy Research Scientific
Computing Center (NERSC). Research supported in part (GMS) by the Division of Materials Science and Engineering, Office of Basic
Energy Science,
U.S. Department of Energy. We gratefully acknowledge helpful discussions with M. S. S. Brooks.

\end{document}